\documentclass{article}
\usepackage{ijcai18}

\usepackage{times}
\usepackage{xcolor}
\usepackage{soul}
\usepackage[utf8]{inputenc}
\usepackage[small]{caption}

\usepackage{graphicx}
\usepackage{url}
\usepackage{booktabs} 
\usepackage{subfig}
\usepackage[para,online,flushleft,normal]{threeparttable}
\usepackage{algorithm}
\usepackage{algorithmic}
\usepackage{amsmath}
\usepackage{amsfonts}
\usepackage{amssymb}

\usepackage{flushend}

\newcommand{\nop}[1]{}

\newcommand{\etc}{{etc.}}

\newtheorem{definition}{Definition}

\title{Meta-Path Constrained Random Walk Inference \\for Large-Scale Heterogeneous Information Networks}

\author{Chenguang Wang \\
        Amazon Web Services \\
        \texttt{chgwang@amazon.com}}

\begin{document}

\maketitle

\begin{abstract}
Heterogeneous information network (HIN) has shown its power of modeling real world data as a multi-typed entity-relation graph. Meta-path is the key contributor to this power since it enables inference by capturing the proximities between entities via rich semantic links. Previous HIN studies ask users to provide either 1) the meta-path(s) directly or 2) biased examples to generate the meta-path(s). However, lots of HINs (e.g., YAGO2 and Freebase) have rich schema consisting of a sophisticated and large number of types of entities and relations. It is impractical for users to provide the meta-path(s) to support the large scale inference, and biased examples will result in incorrect meta-path based inference, thus limit the power of the meta-path. In this paper, we propose a meta-path constrained inference framework to further release the ability of the meta-path, by efficiently learning the HIN inference patterns via a carefully designed tree structure; and performing unbiased random walk inference with little user guidance. The experiment results on YAGO2 and DBLP datasets show the state-of-the-art performance of the meta-path constrained inference framework.
\end{abstract}

\section{Introduction}
Heterogeneous information networks (HINs) consisting of multi-typed entities and relations have been extensively studied recently, and shown the power against many other state-of-the-art data models in lots of real world applications, such as link prediction~\cite{taskar2003link,lu2011link} and classification~\cite{ji2010graph,wang2016knowledgekernel}, etc. The main reason behind the success of the HIN studies is the concept of meta-path~\cite{sun2011pathsim}. Meta-path is a sequence of consecutive entity types and relation types capturing semantic proximity between entities in HINs, thus provides the opportunities to deeply understand the data. 

Most HIN tasks can be formalized as inference with meta-paths problems in HINs. The goal of HIN inference is to find the best assignment to the output variables according to the given model and input instances. HIN inference particularly leverages the proximities between the input and output variables provided by the meta-paths, i.e., $\mathbf{y}^{*} = \arg \max_{\mathbf{y}}f(\mbox{\boldmath$\theta$}, \Phi(\mathbf{x},\mathbf{P},\mathbf{y}))$, where $f$ is an HIN model, $\mbox{\boldmath$\theta$}$ is the model parameters, $\Phi$ is a set of representation (i.e., feature) functions based on input instance $\mathbf{x}$, output variables $\mathbf{y}$ and meta-paths $\mathbf{P}$ relevant to $\mathbf{x}$ and $\mathbf{y}$. For example, in HIN based link prediction task, given the linear regression based prediction model $f$ and the feature function set $\Phi$ built upon source entity $\mathbf{s}$ and target entity $\mathbf{t}$ (e.g., $\mathbf{x} = [\mathbf{s}, \mathbf{t}]$), meta-path based link prediction can be seen as finding the best assignment to output variable $\mathbf{y}$ (one-dimension vector, e.g., {\bf [1]} represents link exists, {\bf [0]} represents link doesn't exist), via the meta-paths $\mathbf{P}$ between $\mathbf{x}$ and $\mathbf{y}$. As we can see, the power of the inference in HINs is brought by meta-path(s). The ability of how well we can handle the meta-paths is the key to better inference in HINs.

However, there are two possible limits in the current inference methods in large scale HINs. 

{\bf User needs to provide the meta-path(s).} 
Most of the previous HIN studies ask users or experts to provide meta-path(s) as explicit inference rules to perform relevant tasks in HINs, such as similarity search~\cite{sun2011pathsim}. Since in traditional HINs, such as DBLP, there are only four types of entities, {\sl Paper}, {\sl Venue}, {\sl Author} and {\sl Term} and relevant relation types between the entities. It will be possible for users to provide high-quality meta-paths. However, in large scale HINs, such as YAGO, it consists of a sophisticated and large number of entity types (e.g., millions) and relation types (e.g., hundreds). Users will feel harder to provide meta-paths in such cases. Moreover, multi-order (length larger than one) meta-paths could carry more important information for inference than that of first-order meta-path (length equals to one). Thus it becomes harder to rely on users to provide the relevant meta-paths for inference in large scale HINs. 

{\bf Biased examples lead to incorrect inference.}
To relieve the efforts of users, random walk based inference~\cite{lao2010relational,lao2011random} is proposed for large scale HINs. Existing random walk procedure aims to enumerate the meta-paths within fixed length $l$, then perform jointly inference. Since the time complexity of enumerating the meta-paths grows exponentially with the length of the meta-paths, traditional random walk will be impractical in large scale network. Besides, another potential problem is that the inference performance is very sensitive to $l$. For example, if $l$ is small (e.g., equals to 1), multi-order and meaningful meta-paths will be ignored; if $l$ is large, meaningless and duplicated meta-paths will be generated. 
Meng et al.~\cite{meng2015discovering} recently provides a more general inference framework by requiring users to provide example entity pairs. Then the meta-paths are generated if relevant meta-paths are of high proximity between the entity pairs. 
The algorithm generates meta-paths by considering the local (near) randomly generated negative examples. Since there could be some positive examples introduced by randomly producing negative samples, the negative examples could be biased and noisy towards the inference.
We thus consider an efficient HIN inference framework that could further release the power of meta-path based inference in large scale HINs by relieving the aforementioned issues.


In this paper, we propose a meta-path constrained inference framework for large scale HIN. Since for a particular HIN inference task, we may only care about a (or a set of) relevant inference rules, then the parts of the network only distantly connected to inference goals are likely to have a small influence. Intuitively, we consider the HIN inference as graph random walk inference with constraints. We carefully design an efficiency-optimized meta-path tree data structure to constrain the random walk to follow the tree structure. Compared to the running time of the original random walk, the proposed meta-path constrained random walk can improve efficiency by mostly two orders of magnitude. The proposed inference scheme is initialized with several sample pairs to capture the inference target. Then based on the meta-path based proximities in the samples, the scheme iterates random walk inference in the meta-path tree until it becomes convergence. Notice that the inference method is weakly supervised. The supervision information complies with the intuition that the inference is only relevant to parts of the HINs.


The main contributions of this paper can be highlighted as below:
\begin{itemize}
\item We study the problem of large scale HIN inference, which is important and has broad applications.
\item We propose a meta-path constrained inference framework, where many of the HIN tasks can be unified under this framework. In particular, we propose a meta-path tree constrained random walk inference method, which is weakly supervised and efficient to model the inference targets.
\item We conduct experiments on two large scale HIN. Our proposed inference method has demonstrated its effectiveness and efficiency compared to the state-of-the-arts inference methods on typical HIN tasks (link prediction and similarity search).
\end{itemize}

The rest of this paper is organized as below. We first introduce the HIN inference framework in Section~\ref{sec:pd}. Next in Section~\ref{sec:ap}, we present HIN inference as meta-path tree constrained random walk inference. The experimental results are shown in Section~\ref{sec:ex}. We finally discuss the related work and conclude in Section~\ref{sec:rw} and Section~\ref{sec:co} respectively.

\section{Problem Definition}
\label{sec:pd}
In this section, we introduce a general HIN inference framework that could specifically take meta-paths in HIN into consideration, and perform large scale inference according particular task. Before that, some basic concepts of HIN are introduced as below.

\subsection{Heterogeneous Information Network}
\begin{figure}[h]
\centering
\includegraphics[width=0.40\textwidth]{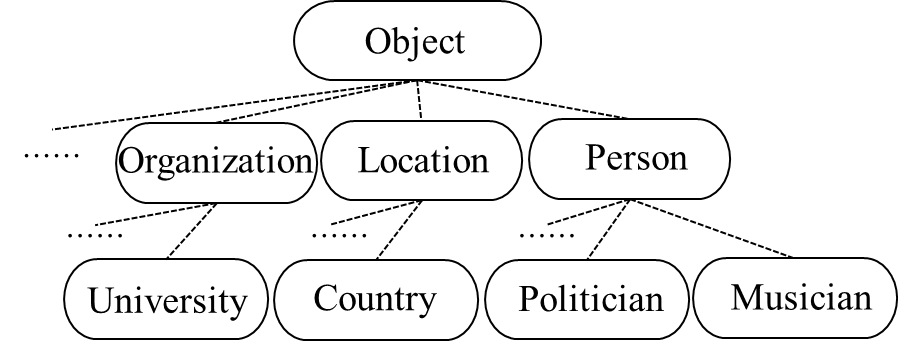}
\caption{Hierarchy of entity types.}
\label{fig:concepthierarchy}
\end{figure}

\textit
{
\begin{definition}
\label{def:hin}
A \textbf{heterogeneous information network} (HIN) is a graph ${\mathcal G} = ({\mathcal V}, {\mathcal E})$ with an entity type mapping $\phi$: ${\mathcal V} \to \mathcal A$ and a relation type mapping $\psi$: ${\mathcal E} \to \mathcal R$, where ${\mathcal V}$ denotes the entity set, ${\mathcal E}$ denotes the link set, $\mathcal A$ denotes the entity type set, and $\mathcal R$ denotes the relation type set, and the number of entity types $|\mathcal A|>1$ or the number of relation types $|\mathcal R|>1$.
\end{definition}
}
Notice that, in large scale HINs, such as YAGO, the relation type mapping is an one-to-one mapping, while the entity type mapping could be one-to-N mapping. For example, a specific triplet (Larry Page, {\it alumniOf}, Stanford) in YAGO, the relation type mapping $\psi(<Larry Page, Stanford>) = \{{\it alumniOf}\}$, while the entity mapping $\phi(Stanford) = \{{\it Organization, University}\}$. The reason why an entity could be mapping to multiple entity types in YAGO or Freebase is that, the entities types are often organized in a hierarchical manner. For example, as shown in Figure~\ref{fig:concepthierarchy}, {\it University} is a subtype of {\it Organization}, {\it Politician} is a subtype of {\it Person}. All the types or attributes share a common root, called {\it Object}. The hierarchy of the entity organization raises another challenge on how to infer useful information in HINs.

\textit
{
\begin{definition}
Given an HIN ${\mathcal G} = ({\mathcal V}, {\mathcal E})$ with the entity type mapping $\phi$: ${\mathcal V} \to \mathcal A$ and the relation type mapping $\psi$: $\mathcal E \to \mathcal R$, the \textbf{network schema} for network $G$, denoted as $\mathcal T_{\mathcal G} = (\mathcal A, \mathcal R)$, is a graph with nodes as entity types from $\mathcal A$ and edges as relation types from $\mathcal R$.
\end{definition}
}

The network schema provides a high-level description of a given heterogeneous information network.
Another important concept, meta-path~\cite{Yizhou11}, is proposed to systematically define relations between entities at the schema level.
\textit
{
\begin{definition}
A \textbf{meta-path} $\mathcal P$ is a path defined on the graph of network schema $\mathcal{T}_G = (\mathcal A, \mathcal R)$, and is denoted in the form of ${\it A_1  \xrightarrow{R_1} A_2 \xrightarrow{R_2}  \dots \xrightarrow{R_L} A_{L+1}}$, which defines a composite relation $R = R_1 \cdot R_2 \cdot \ldots \cdot R_L$ between types $A_1$ and $A_{L+1}$, where $\cdot$ denotes relation composition operator, and $L$ is the length of $\mathcal P$. When $L = 1$, we specifically call it as {\it first-order meta-path}; when $L > 1$, we call it as {\it multi-order meta-path}.
\label{def:mp}
\end{definition}
}


We say a path $p = (v_1-v_2- \ldots -v_{L+1})$ between $v_1$ and $v_{L+1}$ in network $\mathcal{G}$ follows the meta-path $\mathcal P$, if $\forall l, \phi(v_l) = A_l$ and each edge
$e_l = \langle v_l,v_{l+1} \rangle$ belongs to each relation type $R_l$ in $\mathcal P$.
We call these paths as {\it path instances} of $\mathcal P$, denoted as $p \in \mathcal P$. $R_l^{-1}$ represents the reverse order of relation $R_l$. For example, in the YAGO network, the composite relation {\it two Person co-founded an Organization} can be described as meta-path
{\small $\mathcal P$ = {\it Person} $\xrightarrow{\rm found}$ {\it Organization} $\xrightarrow{\rm found^{-1}}$ {\it Person}}. A path instance of $\mathcal P$ is {\small $p$ = Larry Page $\xrightarrow{\rm found}$ {\it Google} $\xrightarrow{\rm found^{-1}}$ Sergey Brin}.

\subsection{Heterogeneous Information Network Inference Framework}

Meta-paths carry rich information about the semantic relationships between entities, thus capture subtle proximities in HINs via several meta-path based similarity measures, such as {\sl Path Count}~\cite{sun2011pathsim}, {\sl Random Walk}~\cite{lao2010relational}, and {\sl Pathsim}~\cite{sun2011pathsim}. The proximities are very important and useful for inference problems, such as link prediction. Assume that direct links (first-order meta-paths) are missing between two entities with types {\it Person} and {\it Profession}, if there is a multi-order meta-path {\small {\it Person} $\xrightarrow{\rm workAt}$ {\it Organization} $\xrightarrow{\rm employ}$ {\it Profession}} between the entities, and the meta-path based similarity (e.g., {\sl Path Count}) of two entities is large (i.e., number of path instances between two entities satisfying the meta-path is large), we will have higher confidence in inferring the missing link between the entities based on the multi-order meta-path.

Traditional inference framework in machine learning~\cite{ChangRR12} aims to model relevant inference problems as stochastic processes involving both output variables and input or observed variables. The framework mainly includes a model parameter vector $\mbox{\boldmath$\theta$}$, corresponding to a set of representation or feature functions $\{\Phi\}$. For an input instance $\mathbf{x}$ and an output assignment $\mathbf{y}$, the ``score'' of the instance can be expressed as a model function $f$ with the parameter vector and representation functions: score = $f(\mbox{\boldmath$\theta$}, \Phi(\mathbf{x},\mathbf{y}))$. When the model is evaluated on test instance $\mathbf{x}$, the inference framework aims to find the best assignment to the output variables,
\begin{equation}
\label{eq:iff}
\mathbf{y}^{*} = \arg \max_{\mathbf{y}}f(\mbox{\boldmath$\theta$}, \Phi(\mathbf{x},\mathbf{y}))
\end{equation} 
Notice that a representation function $\phi \in \Phi$ usually focuses on producing homogeneous flat features and ignores the link or structure information in both input variables $\mathbf{x}$ and output variables $\mathbf{y}$.
As we know, the multi-typed links (meta-paths) is very important for HIN inference. However, the framework doesn't model meta-path information; besides it is not trivial to efficiently incorporate the extra information provided by meta-paths into the framework. 
We therefore formally define the HIN inference framework as below.
\begin{definition}
\textbf{Heterogeneous Information Network Inference (HINI)} aims to enable inference with meta-paths in large scale HINs. To be more specific, HINI aims to infer the best assignment to the output variables given an HIN model with meta-paths. HINI is formalized as:
\begin{equation}
\label{eq:hini}
\mathbf{y}^{*} = \arg \max_{\mathbf{y}}f(\mbox{\boldmath$\theta$}, \Phi(\mathbf{x}, \mathbf{P}, \mathbf{y}))
\end{equation} 
where $f$ is the model or score function and $\mbox{\boldmath$\theta$}$ is the model parameters. Different from Eq.~\ref{eq:iff}, $\Phi$ aims to leverage the proximities carried by the meta-paths $\mathbf{P}$ regarding to the input instance $\mathbf{x}$ or output variable $\mathbf{y}$.
\end{definition}

HINI (Eq.~\ref{eq:hini}) could support many mining tasks in HINs, such as link prediction and similarity search in HINs, as we will see later. If meta-paths set $\mathbf{P}$ is empty. HINI will degenerate to traditional inference as shown in Eq.~\ref{eq:iff}.
We find there are mainly two challenges in HINI: 1) how to efficiently generate useful meta-paths $\mathbf{P}$ from large-scale HINs consisting of millions of entities and billions of relations? And 2) how to model the meta-path based proximities to improve the representation power of $\Phi$? In next section, we will describe our proposed inference method that could efficiently generate meta-paths as well as compute meta-path based similarity simultaneously.

\section{HIN Random Walk Inference with Meta-Path Dependency Tree Search}
\label{sec:ap}

In this section, we first introduce meta-path constrained HIN random walk inference with weak supervision, then talk about how to leverage the supervision to conduct efficient HIN random walk inference via a carefully designed data structure.

\subsection{Weakly Supervised HIN Random Walk Inference}
As aforementioned, meta-paths are very important since they infer important semantic relationships between entities in HINs, thus capture semantic proximities of entities, which is very useful for HIN inference as shown in Eq.~\ref{eq:hini}. Most of existing inference methods are focusing on enumerating the meta-paths within a fixed length in the full underlying network~\cite{lao2010relational}. However, this solution is impractical in large scale HINs, since it has been proven that the number of possible meta-paths grows exponentially with the length of a meta-path. As we find, for a particular inference task, it is not necessary to do inference in the full HIN, since only a part of the network or meta-paths relevant to the inference. To copy with these challenges, we propose a meta-path constrained random walk method to infer with weak human supervision in HINs. Weak supervision here provides guidance to the random walk process, together with the meta-path generation process by pruning the searching space. Our method aims to copy with the two HINI challenges in the previous section.

Given an HIN ${\mathcal G} = ({\mathcal V}, {\mathcal E})$, similar to~\cite{meng2015discovering}, we ask users to provide example entity pairs $\mathbf{\Lambda} = \{(s_i, t_i) | i = 1, \cdots, N\}$ as supervision to imply meta-paths. Formally, we are aiming to find a meta-paths set $\mathbf{P}_{\mathbf{\Lambda} } = \{\mathcal{P}_i | i = 1, \cdots, M\}$ that could infer high proximities between entities in $\mathbf{\Lambda} $. An efficient way to generate $\mathbf{P}_{\mathbf{\Lambda} }$ given $\mathbf{\Lambda} $ will be described in next section.

Now assume when we have $\mathbf{\Lambda} $, we also get $\mathbf{P}_{\mathbf{\Lambda} }$. For a meta-path $\mathcal{P} = {\it A_1  \xrightarrow{R_1} A_2 \xrightarrow{R_2}  \dots A_i \xrightarrow{R_i} \dots \xrightarrow{R_L} A_{L+1}} \in \mathbf{P}_{\mathbf{\Lambda} }$, we define the following meta-path constrained random walk starting from $s$ and reaching at $s$ following only path instances $p \in \mathcal{P}$. It defines a distribution $f(s,t | \mathcal{P})$ recursively as below.
\small{
\begin{align}
\label{eq:onerw}
f(s,t | \mathcal{P}_{1,\cdots,L}) = \sum_{t' \in \chi(A_{i+1}|e_i; R_i)}f(s,t' | \mathcal{P}_{1,\cdots,L-1}) \nonumber \\ \cdot \frac{1}{|\chi(A_{i+1}|e_i; R_i)|},
\end{align}
}
\begin{equation}
\text{s.t. }f(s,t | \mathcal{P}) \text{ = 1, if s = t}; \nonumber \\
\text{otherwise, }f(s,t | \mathcal{P}) = 0, \nonumber
\end{equation}
where $\chi(A_{i+1}|e_i; R_i)$ indicates the entity set where each entity $e_i$ can be linked via relation type $R_i$ to at least one entity with type $A_{i+1}$. $\frac{1}{|\chi(A_{i+1}|e_i; R_i)|}$ means a one-step random walk starting from an entity $t' \in \chi(A_{i+1}|e_i; R_i)$ via relation type $R_i$.

For example, consider a path instance from $s$ to $t$ following a meta-path {\it Person} $\xrightarrow{\rm workAt}$ {\it Organization} $\xrightarrow{\rm employ}$ {\it Profession}. Suppose a random walk starts at an entity $s$ (e.g., $s$=Larry Page). If $\mathbf{Org}$ is the set of {\it Organization}s in the HIN that Larry Page has worked at, after one step, the walker will have probability $\frac{1}{|\mathbf{Org}|}$ of being at any entity $e \in \mathbf{Org}$ (e.g., $e$= Google). Similarly, if $\mathbf{Pro}$ is the set of {\it Profession}s in the HIN that Google has employed, the walker will have probability of $\frac{1}{|\mathbf{Org}|\cdot|\mathbf{Pro}|}$ being at any entity $t \in \mathbf{Pro}$ (e.g., $t$ = Computer Science).
It is useful that proximity provided by meta-path constrained random walk infers the prior probability of $t$ being the {\it Profession} for {\it Person} $s$.

To be more general, we then propose a linear model to combine the single meta-path constrained random walk scores. i.e., for each meta-path $\mathcal{P} \in \mathbf{P}_{\mathbf{\Lambda}}$, the {\it HIN inference with joint random walk model} (HINI-JRW) is formalized as below.
\begin{align}
\label{eq:jointmodel}
f(s,t | \mathbf{P}_{\mathbf{\Lambda}}; ) = \theta_1f(s,t | \mathcal{P}_1) + \theta_2f(s,t | \mathcal{P}_2)  + \cdots  \nonumber \\
+ \theta_if(s,t | \mathcal{P}_i) + \cdots + \theta_Mf(s,t | \mathcal{P}_M),
\end{align}
where is the model parameter vector, each element means the weight or importance of certain meta-path based inference score. The parameter vector can either be explicitly set by users or implicitly learned according different HIN inference tasks. By tuning the parameters, we can avoid the bias induced by certain meta-path(s) and ensure the model's robustness and stability.

Now let's revisit the relationship between the joint HIN inference model (Eq.~\ref{eq:jointmodel}) and HINI framework (Eq.~\ref{eq:hini}). In short, the input instance of HINI $\mathbf{x} = [s,t]$, the output variable of HINI $\mathbf{y}$ is the random walk based probabilities or proximities, and meta-paths set $\mathbf{P}$ of HINI equals to $\mathbf{P}_{\mathbf{\Lambda}}$. Similarly, other inference models can also be unified into HINI framework.

\subsection{Efficient Inference via Meta-Path Dependency Tree Search}
To copy with the main challenge of large scale HIN inference, i.e., to do efficient inference, we carefully design a tree structure that significantly accelerate the above meta-path constrained random walk inference in HINs. More importantly, the new tree structure enables doing inference through model $f$ in Eq.~\ref{eq:jointmodel} and automatically generate meta-path based inference rules $\mathbf{P}_{\mathbf{\Lambda}}$ given the user provided samples $\mathbf{\Lambda}$ simultaneously.

Even we do not need to enumerate the meta-paths in an HIN, it is still intractable to find the optimal meta-path set $\mathbf{P}_{\mathbf{\Lambda}}$ given the examples pairs in $\mathbf{\Lambda}$~\cite{van1994minimum}. Selecting relevant meta-paths is NP-hard even when some of the path instances are given. It has been shown as an NP-hard problem~\cite{amaldi1998approximability,van1994minimum}. Inspired by forward selection algorithms~\cite{koller1996toward}, we introduce meta-path dependency tree search algorithm, and enable efficient inference of HINI-JRW via doing search in the meta-path dependency tree. Since we do not know the relevant paths beforehand, thus $\mathbf{P}_{\mathbf{\Lambda}}$ is empty. The meta-path dependency tree search algorithm is thus used to insert meta-paths into $\mathbf{P}_{\mathbf{\Lambda}}$ and compute the inference score while searching. There are three differences compared to the method proposed in~\cite{meng2015discovering}: 1) our tree search algorithm doesn't leverage random generated negative examples, which could induce biased inference; 2) our algorithm doesn't require iterative process of the algorithm to guide the tree search, which is more efficient; and 3) we do not set hard threshold to terminate the algorithm, which is very sensitive and could prevent the algorithm from generating more meaningful meta-paths.

\begin{figure}[htbp]
\centering
{
\includegraphics[scale=0.4]{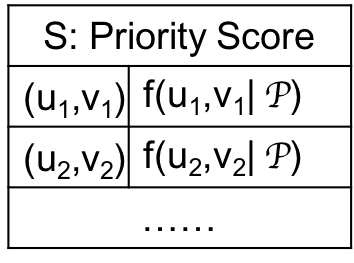}
}
\caption{Meta-path dependency tree node structure.}
\label{fig:nd}
\end{figure}

Let's first introduce the meta-path dependency tree structure. Each tree edge is annotated
with an relation type, and each tree node represents a list of entities pairs
with their HINI-JRW scores and a priority score $S$. The node structure is shown in Figure~\ref{fig:nd}. The node stores tuple in the form $<(s,t), f(s,t | \mathcal{P}_i)>$, where $(s,t)$ represents a path instance in the graph by its starting and current graph entities, respectively; $\mathcal{P}_i$ is the meta-path starting from $s$ to $t$. The edges of the
tree are relation types in the HIN. $f(s,t | \mathcal{P}_i)$ is the RW score
for this meta-path according to Eq.~\ref{eq:onerw}. The priority score $S$ determines which is the next tree node to search. We compute the value of $S$ as follows:
\begin{equation}
\label{eq:priority_score}
S=\left\{
\begin{aligned}
&\frac{\sum_{u}\frac{\max(r_u)\cdot\sum_{v}f(u,v|\mathcal{P})}{c(u)}}{\sum_{v}\max(r_u)}\cdot\beta^{L}       &\nexists(u,v) \in \mathbf{\Lambda}\\
&\frac{\sum_{u}\frac{\max(r_u)\cdot\sum_{v}f(u,v|\mathcal{P})}{c(u)}}{\sum_{v}\max(r_u)}\cdot\beta^{L} + 1   &\exists(u,v)\in\mathbf{\Lambda}\\
\end{aligned}
\right.
\end{equation}


where $(u,v)$ is the entity pairs in the tree node, $\max(r_u)$ is the maximum weight among entity pairs starting with $u$, and $c(u)$ is the number of example pairs starting with $u$. From Eq.~\ref{eq:priority_score}, we can see that the priority score $S$ is defined as a weighed combination of random walk score referred as Eq.~\ref{eq:onerw} of the given entity pair in the tree node. This means that if the entity pair in the tree node exhibit higher proximity along the corresponding meta-path, they will have higher probability to be more similar along the new meta-paths generated through the searching process. This is the reason why the search algorithm will be introduced soon guide its search as random walker to the tree node with larger $S$. Because of this, the priority score proposed in~\cite{meng2015discovering} will have the possibility to generate meta-paths that do not contain example entity pairs in $\mathbf{\Lambda}$. This will guide search to find leaf tree node with the largest $S$ among all the leaf nodes, which will lead to no meta-path generated during the search, thus suffer from relatively low efficiency. To avoid this, considering the value of $S$ is normally smaller than one, this slight modification will ensure the target node $t$ could be reached through the search, and generate the meta-paths in a more efficient way. In addition, in order to avoid generating meta-path with infinite length, we follow~\cite{meng2015discovering} to add a decay factor $\beta$ ranging from 0 to 1.

Through searching the meta-path dependency tree, inference and meta-path generation in HINs can be done at the same time.
We present the details of our proposed meta-path dependency tree search (MPDTS) in Algorithm~\ref{alg:MPT}. In short, we search the tree by moving to out-neighbor nodes on the graph until a meta-path can be found or the graph is completely traversed. We first target
the tree node with largest priority and examine whether its tuples are example entity pairs. If so, we store random walk scores computed by Eq.~\ref{eq:onerw} in the corresponding tree node, and add the meta-paths into $\mathbf{P}_{\mathbf{\Lambda}}$. If no example pairs are encountered, then we extend each entity pair by moving to an out-neighbour. We insert this
new pair with its random walk score to a child node, and compute its priority score. Notice that, the generated meta-path only contains relation types, similar to~\cite{meng2015discovering}, we also leverage the Lowest Common Ancestor (LCS) of type hierarchy of the entities in the HIN to fill the entity types in the generated meta-path $\mathcal{P}$, to form the complete meta-paths.

\begin{algorithm}
\caption{\label{alg:MPT}MPDTS$(\mathcal{G},\mathbf{\Lambda},\mathcal{T})$}
\begin{algorithmic}[1]
\REQUIRE{an HIN $\mathcal{G}(\mathcal{V},\mathcal{E})$, example pairs $\mathbf{\Lambda}$, example pairs weights $\mathbf{r}$, Meta-Path Dependency Tree $\mathcal{T}$}
\ENSURE{Meta-path $\mathcal{P}$, random walk scores vector $\mathbf{y}$ of the corresponding meta-path $\mathcal{P}$}
\WHILE {$\mathcal{T}$ is not fully traversed}
    \STATE $Q \leftarrow$ the largest $S$ score node in $\mathcal{T}$'s child node set $\mathcal{L}$;
    \STATE $\mathbf{y}\leftarrow\{0,\ldots,0\}$;
    \FOR{Each entity pair $(u,v) \in Q$}
        \IF{$(u,v) \in \mathbf{\Lambda}$}
            \STATE $\mathbf{y}(u,v)\leftarrow f(u,v|\mathcal{P})$ computed based on Eq.~\ref{eq:onerw};
        \ENDIF
    \ENDFOR
    \IF{$\mathbf{y}$ is not empty}
        \STATE $\mathcal{P}\leftarrow$ the path from root node to $Q$;
        \STATE Remove $Q$ from $\mathcal{L}$;
        \STATE break;
    \ELSE
        \FOR{Each entity pair $(u,v) \in Q$}
            \FOR{Each out-neighbour $w$ of $v$ in $\mathcal{G}$}
                \STATE $R\leftarrow$relation type of edge from $v$ to $w$;
                \IF{There is no edge of type $R$ from $Q$ to child node $Z$}
                    \STATE Create child node $Z$ for $Q$;
                    \STATE Insert child node $Z$ into $\mathcal{L}$;
                \ENDIF
                \STATE $\mathcal{P}\leftarrow$the path from root node to $Z$;
                \STATE Insert a new tuple $\langle (u,v),f(u,v|\mathcal{P})\rangle$ into $Z$, where $f(u,v|\mathcal{P})$ is computed based on Eq.~\ref{eq:onerw};
                \STATE Compute the priority score$S$ of $Z$ based on Eq.~\ref{eq:priority_score};
                \STATE Remove $Q$ from $\mathcal{L}$;
            \ENDFOR
        \ENDFOR
    \ENDIF
\ENDWHILE
\RETURN $\mathcal{P}, \mathbf{y}$
\end{algorithmic}
\end{algorithm}

Finally, we obtain the joint random walk scores/probabilities along the meta-paths $\mathbf{P}_{\mathbf{\Lambda}}$ according to HINI-JRW (Eq.~\ref{eq:jointmodel}). The joint version of MPDTS (Algorithm~\ref{alg:MPT}) is shown as JMPDTS in Algorithm~\ref{alg:JMPT}. We can simply obtain the joint random walk score for certain entity pair $(s,t) \in \mathbf{\Lambda}$ by summing over the rows of $\mathbf{Y}$, since each entry in the row means a random walk score from $s$ to $t$ following one meta-path. By doing so, we have done inference and automatic meta-path generation in large scale HINs, which copies with the two challenges of HINI framework.

\begin{algorithm}
\caption{\label{alg:JMPT}JMPDTS$(\mathcal{G},\mathbf{\Lambda}, \mathbf{r})$}
\begin{algorithmic}[1]
\REQUIRE{an HIN $\mathcal{G}(\mathcal{V},\mathcal{E})$, example pairs $\mathbf{\Lambda}$, example pairs weights $\mathbf{r}$}
\ENSURE{Meta-path set $\mathbf{P}_{\mathbf{\Lambda}}$ with size $M$, random walk scores matrix $\mathbf{Y}$}
\STATE $m\leftarrow 0$; $\mathbf{P}_{\mathbf{\Lambda}}\leftarrow\varnothing$; $\mathbf{Y}\leftarrow\varnothing$
\WHILE {True}
    \IF{$\mathcal{T}$ is not fully traversed}
        \STATE $\mathcal{P}_m,\mathbf{y}_m \leftarrow$ MPDTS$(\mathcal{G},\mathbf{\Lambda},\mathbf{r})$;
         \STATE $\mathbf{Y}\leftarrow \mathbf{Y}\bigcup \mathbf{y}_m$;
         \STATE $\mathbf{P}_{\mathbf{\Lambda}}\leftarrow \mathbf{P}_{\mathbf{\Lambda}}\bigcup \mathcal{P}_m$;
         \STATE $m\leftarrow m+1$;
    \ENDIF
\ENDWHILE
\RETURN $\mathbf{P}_{\mathbf{\Lambda}},\mathbf{Y}$
\end{algorithmic}
\end{algorithm}

There are several advantages for using the tree structure for random walk inference we proposed. Firstly, during the search process of the tree, the node to expand is selected by applying supervision provided by user examples and the search space can be reduced. Secondly,during the process, path instances that represent the same meta-paths are gathered in a single tree node, yet avoid duplicate calculation. Thirdly, the whole meta-path dependency tree structure is preserved in the memory and can be reused in the whole iterative process while traversing the tree.

\section{Experiments} 
\label{sec:ex}

In this section, we evaluate our approach using two typical HIN tasks, link prediction and similarity search.

\subsection{HINI for Link Prediction}
In this subsection, we validate our algorithm's efficiency and effectiveness 
by performing link prediction tasks.
We chose link prediction for the evaluation because it provides quantitative way to measure the performance of different methods.

\subsubsection*{Task Description}
For link predication, we propose to use logistic regression model to leverage the random walk score of each meta-path~\cite{lao2010relational}. Formally, Given a relation $R$ and a set of entity pairs $\{(s_i, t_i)\}$, we can construct a training dataset $\mathbf{D} = {(\mathbf{x}_i, r_i)}$, where $\mathbf{x}_i$ is a vector of all the meta-path based features for the pair $(s_i, t_i)$—i.e., the j-th component of $\mathbf{x}$ is $f(s_i, t_i | \mathcal{P}_j)$, and $r_i$ indicates whether $R(s_i, t_i)$ is true. Parameter $\theta$ is estimated by maximizing the regularized objective function proposed in~\cite{lao2010relational}. Then the link predication model is defined as below:
\begin{equation}
p(r_i = 1 | \mathbf{x}_i; \theta) = \frac{exp(\theta^T \mathbf{x}_i)}{1+exp(\theta^T \mathbf{x}_i)}.
\end{equation}

\subsubsection*{Dataset}
We perform link prediction experiments on a representative dataset for large scale HIN: YAGO.

{\bf  YAGO:} YAGO\footnote{\url{http://www.mpi-inf.mpg.de/departments/databases-and-information-systems/research/yago-naga/yago/}} is a semantic knowledge base, derived from Wikipedia, WordNet and GeoNames. Currently, YAGO2 has knowledge of more than 10 million entities (like persons, organizations, cities, \etc) and contains more than 120 million facts about these entities. It contains 350,000 entity types organized in type hierarchy, and 100 relation types.

\subsubsection*{Effectiveness Study}

For a certain type of link in YAGO, for instance {\sl citizenOf}, we
remove all such links and try to predict them with the above logistic regression model that leverages the random walk scores as features. We randomly select a number of pairs of entity pairs with the according relation labels as training data, and validate the model using a test set of an equal number of pairs. We compared our link prediction model with PCRW~\cite{lao2010relational} based models which generate paths of finite length in {1,2,3,4}. The PCRW models also use the logistic regression model to combine these meta-paths. Besides, we also compare our model with the state-of-the-art FSPG based prediction model in~\cite{meng2015discovering}.
Following~\cite{meng2015discovering}, we set $\beta$ as 0.6 in meta-path dependency tree to avoid producing meta-paths with infinite length.
We use Area Under ROC Curve (AUC) as the evaluation measure. AUC calculate the area under Receiver Operating Characteristics (ROC) curve. The x-axis of ROC is false positive rate, and the y-axis of ROC represents true positive rate. Thus a large AUC value, a large accuracy in predication.

Table~\ref{tab:lpp} presents the results for link prediction 
for three types of links: {\sl citizenOf} and{\sl advisorOf} in YAGO.
For each of these links, we generated 100
training and 100 test pairs, as described above.
The result shows that fixed-length PCRW suffers from several
issues. When the maximum length is too small (1 or 2), meta-paths
cannot connect example pairs, and as such the model is not better
than a random guess and the model will have low recall. When
the maximum length is too big, the model introduces too many
meta-paths. For length 3, there are 135 meta-paths, and over 2,000
for length 4. Notice that our method outperforms FSPG, the reason is that FSPG incorporates randomly generated negative examples which may lead to biased inference, while our method does not use.

\begin{table}[htbp]
\centering
\caption{Comparison of link prediction performance (AUC)}
\label{tab:auc}
\begin{threeparttable}
\begin{tabular}{|c|c|c|}
\hline
Advisor & citizenOf & advisorOf \\ \hline
Our method & {\bf 0.854} & {\bf 0.654} \\ \hline
FSPG & 0.822  &   0.647 \\ \hline
PCRW-1 & 0.594 & 0.498 \\ \hline
PCRW-2 & 0.752 & 0.613 \\ \hline
PCRW-3 & 0.567 & 0.545 \\ \hline
PCRW-4 & 0.525 & 0.569 \\ \hline
\end{tabular}
\end{threeparttable}
\label{tab:lpp}
\end{table}

\begin{table*}[htbp]
\centering
\caption{Meta-path examples generated based on example pairs}
\label{tab:path_show}
\begin{tabular}{|c|}
\hline
Meta-paths \\ \hline
\emph{Venue$\xrightarrow{\text{publishIn}^{-1}}$Paper$\xrightarrow{\text{authorOf}^{-1}}$Author$\xrightarrow{\text{authorOf}}$Paper$\xrightarrow{\text{publishIn}}$Venue} \\ \hline
\emph{Venue$\xrightarrow{\text{publishIn}^{-1}}$Paper$\xrightarrow{\text{cite}}$Paper$\xrightarrow{\text{publishIn}}$Venue} \\ \hline
\emph{Venue$\xrightarrow{\text{publishIn}^{-1}}$Paper$\xrightarrow{\text{cite}^{-1}}$Paper$\xrightarrow{\text{publishIn}}$Venue} \\ \hline
\end{tabular}
\end{table*}

\begin{table*}[htbp]
\centering
\caption{Case study on similarity search results}
\label{tab:path_show1}
\begin{tabular}{|c|c|c|c|}
\hline
Ranking (Query) & KDD & ACL & VLDB  \\ \hline
1  & ICML & COLING & SIGMOD \\ \hline
2  & SIGMOD & Computational Linguistics & ICDE \\ \hline
3  & ICDM & NAACL & TODS \\ \hline
4  & CIKM & EACL & TKDE \\ \hline
5  & VLDB & LREC & PODS \\ \hline
6  & SIGIR & EMNLP & VLDB Journal\\ \hline
7  & WWW & INTERSPEECH  & EDBT \\ \hline
8  & Machine Learning & SIGIR & PVLDB \\ \hline
9  & TKDE & IJCAI & CIKM \\ \hline
10  & ICDE & AAAI & IS \\ \hline
\end{tabular}
\end{table*}

Our model is clearly better in predicting the links, and it generates
only a limited number of meta-paths. For instance, the {\sl advisorOf}
link in YAGO has a model of only 13 meta-paths. Moreover, these
meta-paths are highly relevant and serve as good explanation of the
proximity links. For example, we find the meta-path illustrating
the fact that a person is a strong influencer to another person is the
best predictor of {\sl advisorOf}, but other, longer, paths are also highly
relevant. Compared with PCRW with maximum length 2, it has
higher recall because it also detected longer important meta-paths,
for instance, {\it Person} $\xrightarrow{\rm win}$ {\it Award} $\xrightarrow{\rm win^{-1}}$ {\it Person}. 
Knowledge bases, such as YAGO, often suffer from incompletion problem.
We thus can use these multi-order meta-paths as inference rules to predict the direct links between entities.

Whereas considering direct links is widely done when querying
large scale networks, multi-order meta-paths can improve query
result accuracy. For instance, in {\sl advisorOf} prediction, only
one type of direct links between people and country exists, namely, 
{\sl influencer}. By leveraging multi-order meta-paths, it will provide us a better chance to still perform good link prediction when the direct links are missing. The result shows the power of our HINI framework in doing large scale inference in HINs.

\subsubsection*{Efficiency Study}
\begin{figure}[htbp]
\centering
{
\includegraphics[scale=0.4]{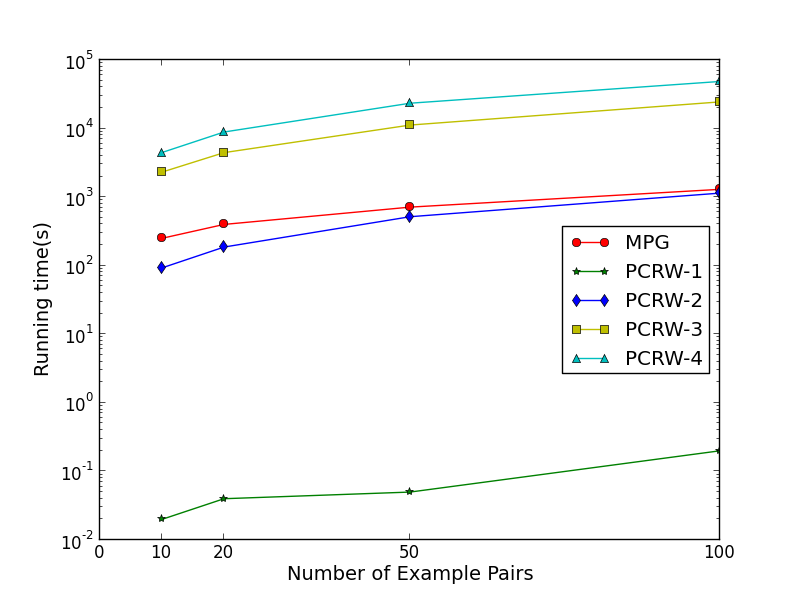}
}
\caption{Running time of our method (MPG)}
\label{fig:running_time}
\end{figure}

Figure~\ref{fig:running_time} presents the running time of our method compared to PCRW
models with fixed length, and when varying the number of example
pairs given as input. It can be observed that generally, the algorithm
running time increases sub-linearly in the number of example pairs.
The increase is due to the PCRW random walks which need to be
performed concurrently for each example pair, but the number of
meta-paths in the model does not increase at the same rate.
In particular, the algorithm performs better than models of paths longer
than 2 by a factor of up to two orders of magnitude. However, the
models of short path length have limited predictive power, despite
their better running time. In comparison, our algorithm is capable of
generating long and meaning meta-paths and performing efficient inference.

\subsection{HINI for Similarity Search}
In this subsection, we mainly focus on user study of our HINI framework via a typical HIN task, i.e., similarity search, to get a better understanding of why our method is effective.
\subsubsection*{Task Description}
In general, similarity search aims to find similar entities given a query entity in HINs. Formally, similarity search can be regarded as random walk, but implemented by commuting matrix manipulation for a meta-path, to generate the similarity score of each targeting entity. The community matrix is defined as below.
\textit
{
\begin{definition}
\textbf{Commuting matrix.}
Given a network ${\mathcal G} = ({\mathcal V}, {\mathcal E})$ and its network schema $\mathcal T_{\mathcal G}$, a commuting matrix $ {\bf M}_{\mathcal P}$ for a
meta-path $\mathcal P = {\it (A_1-A_2- \ldots -A_{L+1})}$ is defined as $ {\bf M}_{\mathcal P} = {\bf W}_{A_1A_2} {\bf W}_{A_2A_3}\ldots {\bf W}_{A_{L}A_{L+1}}$, where ${\bf W}_{A_iA_j}$ is the adjacency matrix between types $ A_i$ and $ A_j$. ${\bf M}_{\mathcal P}(i,j)$ represents the number of path instances
between objects $x_i$ and $y_j$, where $\phi(x_i) = A_1$ and $\phi(y_j) = A_{L+1}$, under meta-path $\mathcal P$.
\label{def:cm}
\end{definition}
}

Given $f(s,t | \mathbf{P})$, we can infer:
\begin{equation}
{\bf M}_{\mathbf{P}} = \sum_i \theta_i \cdot {\bf M}_{{\mathcal{P}}_i}.
\end{equation}
Thus in general, if the entry of ${\bf M}_{\mathcal P}$ is large, the similarity between two entities based on meta-path ${\mathcal P}$ will be large.
\subsubsection*{Dataset}

Most of the similarity search research in HINs are evaluated on DBLP dataset, we also use it for our experiments.
{\bf DBLP} is a bibliographic information network which is frequently
used in the study of heterogeneous networks. We use a subset of DBLP used in~\cite{sun2011pathsim} containing scientific papers in four areas: databases,
data mining, artificial intelligence, and information retrieval. The
dataset has four classes of nodes: Paper, Author, Topic, and Venue.
It also has four edge types between different entity types. In totally, the subset contains 14,376 papers, 14,475 authors, 8,920 topics,
and 20 venues. Besides, there are 170,794 links.

\subsubsection*{Case Study of Similarity Search Results}

We select ten groups of similar venues according to different areas in DBLP, and use them as input example pairs to generate the meta-paths. Then we compute the commuting matrices for the meta-paths. When given a query venue, we rank the venues based on the scores in the row of certain commuting matrix for meta-path. By doing this, we could find the most similar venues to the query.

In Table~\ref{tab:path_show} and Table~\ref{tab:path_show1}, we show the generated meta-paths, and the search results based on the meta-paths. From the results, we find that 1) meta-paths carry rich proximity information between the given pairs; and 2) all the search results are very relevant to the query. This shows the insight of why our HIN inference framework is able to handle automatic meta-path generation and effective inference. By providing these two HIN tasks, we can conclude that the proposed inference framework is not limited to use only in these two tasks, it will be effective in other HIN tasks, such as recommendation, classification, etc.
\section{Related Work}
\label{sec:rw}
\subsection{Knowledge Base/Network Inference}
Although there is a great deal of recent research
on extracting knowledge from text~\cite{agichtein2000snowball,etzioni2005unsupervised,snow2006semantic,pantel2006espresso,banko2007open,yates2007textrunner}, much less
progress has been made on the problem of drawing
reliable inferences from this imperfectly extracted
knowledge. In particular, traditional logical
inference methods are too brittle to be used to make
complex inferences from automatically-extracted
knowledge, and probabilistic inference methods~\cite{richardson2006markov} suffer from
scalability problems, since they cannot generate inference rules directly and use the full rule set to perform inference.
Recently, Ni Lao et al.~\cite{lao2010relational} provides a probabilistic way to perform random walk inference, however, it is still costful since it needs to enumerate the network to generate the meta-paths. However, we carefully design a tree structure which would efficiently generate the meta-paths.

\subsection{Meta-Path Generation}
\nop{
The issue of generating good meta-paths
has not been satisfactorily addressed. A simple method is to
first enumerate all possible meta-paths, by traversing the schema
graph [20], and then use the paths as features to train a regression
model that best fits the user-provided example pairs. However, the
cost of generating all path patterns is prohibitive when the number of
node types and edge types is large. Second, the high dimensionality
of the resulting data means that, for each feature and example pair,
we need to generate the similarity measures by performing the
random walks. Moreover, the number
of features is highly likely to introduce noise due to the curse of
dimensionality [6]. On the contrary, our method only generates
relevant meta-paths. The same authors in [20] also suggest to hire
domain experts to define meta-paths. As we have explained, this
may not be feasible for very large HIN. Another problem with this
approach is that the paths defined in this way are global. In our
solution, users can participate in the process of generating meta-paths
by suggesting example pairs.
}
The first approach regarding to automatic meta-path generation is proposed by~\cite{lao2010relational}.
Their solution enumerate all the meta-paths within a fixed length L.
However, it is not clear how L should be set. More importantly, L can
significantly affect the meta-paths generated: (a) if L is large, then
many redundant meta-paths may be returned, leading to curse-of dimensionality
effects; and (b) if L is small, important meta-paths
with length larger than L might be missed. Our experiments have
shown that the running time of the meta-path generation process
grows exponentially with length L. Moreover, the accuracy can
also drop with increase in L. Recent studies also propose efficient meta-path generation algorithms based on localized random walk~\cite{WangSERZH15,WangSLZH15,wang2016knowledgekernel,WangSRZH16}. Recently, Meng et al.~\cite{meng2015discovering} propose to leverage positive and negative examples to generate meta-paths. However, our solution does not leverage the negative examples, which is randomly generated and would lead to biased inference.

\subsection{Link Prediction}
Since the main focus of our work is to generate
meta-paths, we only use link prediction to quantify our advantage
compared with the existing meta-path generation methods.
Compared with~\cite{nickel2012factorizing}, our method predicts
the relationship between different entities, rather than predicting the
types of entities. \cite{nickel2011three} use the factorization of a three-way tensor
to perform relational learning; it only considered simple node types
and is not applicable to our experiment which has both complex
node classes and edge types. Besides, our method is aiming to support efficient link prediction via HINI with random walk.

\subsection{Similarity Search}
Similarity measures have been a hot research topic for years.
They can be categorized broadly into two types: entity similarity measures and relation similarity measures.
Similarity measures, such as SimRank~\cite{jeh2002simrank}, P-Rank~\cite{zhao2009p}, PathSim~\cite{sun2011pathsim}, PCRW~\cite{lao2010relational} and RoleSim~\cite{jin2011axiomatic} capture entity similarity.
Recent studies also focus on similarity search in schema-rich networks~\cite{WangSRWHJZ15,WangSSHSWZ16,WangSLSZ017,WangSLZH18}.
Recent studies on entity similarity also find rules/meta-paths very useful. Path ranking algorithm~\cite{lao2010relational}, rule mining~\cite{galarraga2013amie} and meta-path generation~\cite{meng2015discovering} have demonstrated the effectiveness of using the mined rules or meta-paths for link prediction-like tasks based on entity similarity, while our work is for leveraging the power of efficient HINI random walk inference method to perform similarity search.
\section{Conclusion}
\label{sec:co}
In this paper, we study the problem of large scale HIN inference, which is important and has broad applications (e.g., link predication, similarity search).
We propose a meta-path constrained inference framework, where many of the existing inference methods can be unified under this framework. In particular, we propose a efficiency-optimized meta-path tree constrained random walk inference method for HIN inference, which is weakly supervised to model the inference target and is approximated in time independent of the network size.
We conduct experiments on two large scale HIN, and our proposed inference method has demonstrated its effectiveness and efficiency compared to the state-of-the-arts inference methods on typical HIN tasks (link prediction and similarity search).
The effectiveness of the inference method is not limited to the two tasks, in the future, we plan to apply our method to more real world applications (e.g., NLP tasks~\cite{WangDZZ13,WangCL17,WangACLXX17,wang190409408}).

\bibliographystyle{named}
\bibliography{sigproc}

\end{document}